\documentclass{openjournal}
\usepackage{acronym}
\usepackage{bm}
\usepackage{balance}
\usepackage{graphicx}
\usepackage{amsmath}
\usepackage{hyperref}

\usepackage{xcolor}
\usepackage{textgreek}
\usepackage[utf8]{inputenc}
\usepackage[english]{babel}

\usepackage{hyperref}
\hypersetup{
    unicode, 
    colorlinks=true,
    linkcolor=linkcolor,
    citecolor=linkcolor,
    filecolor=linkcolor,
    urlcolor=linkcolor,
}
\usepackage{color,colortbl}
\definecolor{linkcolor}{rgb}{0.0,0.3,0.5}
\usepackage{tensind}
\tensordelimiter{?}
\DeclareGraphicsExtensions{.bmp,.png,.jpg,.pdf}
\usepackage{verbatim}
\usepackage[normalem]{ulem}
\usepackage{orcidlink}
\usepackage{soul}

\graphicspath{ {./figs/} }

\begin{document}
\title{The Chirp-Mass Ladder: A New Rung Emerges}

\author{Vaibhav Tiwari\orcidlink{0000-0002-1602-4176}}
\email{vaibhavtewari@gmail.com}
\affiliation{Institute of Gravitational Wave Astronomy, School of Physics and Astronomy, University of Birmingham, Edgbaston}
%\affiliation{Your second nice institute / university}

\acrodef{PE}{Parameter Estimation}
\acrodef{MC}{Monte Carlo}
\acrodef{MCMC}{Markov Chain Monte Carlo}
\acrodef{PDF}{Probability Density Function}
\acrodef{NS}{Neutron Star}
\acrodef{BH}{Black Hole}
\acrodef{BBH}[BBH]{binary black hole}
\acrodef{BNS}[BNS]{Binary Neutron Star}
\acrodef{CBC}{Compact Binary Coalesence}
\acrodef{EDF}{empirical distribution function}
\acrodef{CDF}{cumulative distribution function}
\acrodef{SNR}{Signal to Noise Ratio}
\acrodef{GW}[GW]{gravitational wave}

\begin{abstract}
We study the \ac{BBH} population observed through \acp{GW} using 256 events from the latest release of GWTC-5.0. The inferred chirp-mass distribution shows prominent peaks at approximately $7.5M_{\odot}$, $14M_{\odot}$, and $27M_{\odot}$, with subsequent peaks spaced by approximately a factor of two. A parsimonious explanation for this structured distribution is a hierarchical merger scenario, in which the first peak arises from mergers of black holes of stellar origin, and higher-mass peaks arise from repeated mergers. Notably, with the addition of new observations, an intermediate peak near $19M_{\odot}$ emerges. This feature was anticipated in earlier work as a consequence of intergenerational mergers involving second- and third-generation (G) black holes, highlighting the predictive expectations of the hierarchical-merger interpretation. Furthermore, two groups of $1G+2G$ mergers recently reported in separate studies can be understood as distinct rungs---$1G+2G$ and $3G+4G$---within this hierarchical chirp-mass ladder, a unification that describes both spin transitions with a single mechanism. Although we observe expected correlations between mass ratios and spins in multiple events across the mass range, the lack of clear signatures across all rungs invites investigation into the role of hierarchical mergers in shaping the \ac{BBH} population.
\end{abstract}

% Write your keywords here

\maketitle

\section{Introduction} 

The fifth GW catalog released by LVK collaborations~\citep{2015CQGra..32g4001L, 2015CQGra..32b4001A, 2021PTEP.2021eA101A} has almost doubled the number of \acp{BBH} observed through \acp{GW} ~\citep{2026arXiv260527225T}. Despite this rapid expansion, the physical processes governing \ac{BBH} formation and evolution remain incompletely understood. In particular, the origin of the observed mass spectrum, mass ratios, and spin distributions remains debated, with proposed contributions from multiple sub-populations~\citep{2021ApJ...910..152Z, 2022ApJ...941L..39W, 2023arXiv230401288G, 2026PhRvL.137b1403B, 2025PhRvD.112l3054S, 2025ApJ...991...17R, 2026ApJ...996...71H, 2026PhRvD.113d3048B, 2026arXiv260317987R, 2026arXiv260527226T, 2025PhRvD.112b3531A, 2026arXiv260600234P, 2026arXiv260525994G}. These include isolated binary evolution in the galactic field (e.g., \cite{2016Natur.534..512B, 2021A&A...647A.153B, 2022PhR...955....1M, 2023ApJ...948..105V}) and dynamical assembly in dense stellar environments such as globular clusters, young star clusters, and active galactic nuclei~(e.g.,~\cite{2016ApJ...831..187A, 2016PhRvD..93h4029R, 2020ApJ...898...25T, 2020MNRAS.498..495D}). Two proposed sub-populations have recently gained interest as explanations for features of the \ac{BBH} population: those associated with the pair-instability mass gap~\citep{2025PhRvD.112f3040A, 2025arXiv250904151T, 2025arXiv250904637A, 2025arXiv250909123A, 2025arXiv250909876G, 2026arXiv260211282M} and those formed via hierarchical assembly. In the latter, separate groups of \acp{BH} with masses near 9$M_\odot$ and 45$M_\odot$ produce merger remnants that subsequently participate in further coalescences~\citep{2025arXiv251105316T, 2026arXiv260107908P, 2026arXiv260103456F, 2026ApJ...999L..30V, 2026arXiv260407456G}.

Motivated by the observation of four well-placed peaks in the chirp-mass distribution following GWTC-2, we suggested a hierarchical-mergers scenario as a parsimonious explanation for the structure observed in the \ac{BBH} mass distribution~\citep{2021ApJ...913L..19T}. A narrow peak around the chirp mass value 7.5$M_\odot$ followed by a dearth of observations in the region 10--12$M_\odot$, and three more peaks around $14M_\odot$, $27M_\odot$ and $50M_\odot$, neatly followed the approximate doubling of masses expected from a hierarchical merger scenario. In this proposal, the first peak corresponds to \acp{BBH} formed by first-generation~(G) \acp{BH} that are stellar remnants, and higher-mass peaks are dominated by \acp{BBH} formed by the assembly of \acp{BH} that are merger remnants of a lower-generation \acp{BBH}.  Since then, we have followed up on this after the release of GWTC-3 and GWTC-4~\citep{2022ApJ...928..155T, 2025arXiv251025579T}. In this article, we report that the chirp-mass distribution retains these peaks after including GWTC-5.0. In addition, the chirp-mass distribution now shows evidence for an intermediate peak around $\sim$19$M_\odot$, appearing between the established peaks. This peak lies near a chirp mass value where intergenerational mergers are expected to contribute. Notably, a feature in this region had been anticipated in earlier work, underscoring the predictive value of hierarchical-merger interpretations~(Section~4 in~\cite{2022ApJ...928..155T}). The consistency of the mass ladder--specifically the emergence of the 19$M_\odot$ peak--suggests that we may be observing a systematic physical process rather than a stochastic superposition of unrelated formation environments.

However, the lack of clear signatures in mass ratios and spins remains a challenge for the hierarchical merger scenario. Determining whether the variation in mass ratio and spins across the mass distribution reflects multiple sub-populations, a consequence of the limited measurability of these parameters, or something else would help clarify whether hierarchical mergers drive the shaping of the mass distribution. Features in the mass ratio and spins could be washed out by stochastic overlap across multiple independent channels; this appears to be a challenging investigation and would likely require thousands of observations. An immediately actionable step is to investigate whether the limited measurability of parameters, which makes them prone to small systematic effects, leads to apparent absences of expected mass ratios and spins, or to distinct mass-dependent groups of hierarchical mergers, and thus instead reflects parameter degeneracies or parameter-dependent systematics.

This article is organised as follows. In Section~\ref{sec:hm}, we describe the hierarchical merger scenario. In Section~\ref{sec:mchirp_distr}, we discuss the chirp mass distribution of \acp{BBH}. In Section~\ref{sec:hispin}, we examine the subset of high-spin BBHs whose component masses and spins are consistent with hierarchical-merger expectations, highlighting how these events relate to the chirp-mass peaks. In Section~\ref{sec:bias}, we discuss the limited measurability of mass ratios and spins.  In Section~\ref{sec:conclude}, we summarise our results.

\section{The Hierarchical Merger Scenario}
\label{sec:hm}
Previously, we noted that the component masses of high-spin \acp{BBH} cluster around values 8.6$M_\odot$~(1G), 16.3$M_\odot$~(2G), 31.0$M_\odot$~(3G), 59.0$M_\odot$~(4G), $\cdots$~\citep{2025arXiv251025579T}~\footnote{The only free parameter is the starting value. This is chosen to maximise the overlap between the location of inter-- and intra-generation mergers in Table~\ref{tab:loc_peaks_mchirp} and the location of peaks in Figure~\ref{sec:mchirp_distr}. A more flexible methodology can be used, but we don't expect it to meaningfully affect our interpretation within the context of the hierarchical merger scenario, as the starting value is tightly constrained by the location of the first peak. The location of this peak is the most robust among all the features.}. After fixing the location of the first generation, these values increase by a factor of 1.9. This factor accounts for the doubling of masses and applies around 5\% mass loss in \acp{GW} corresponding to low-spin, comparable-mass \acp{BBH}, which will slightly decrease on decreasing the mass ratio and increase~(decrease) on increasing~(decreasing) the aligned spins~\citep{JimenezForteza2017}. The two-component masses of a \ac{BBH} will be distributed around these values, irrespective of whether this \ac{BBH} is intragenerational~(both \acp{BH} are from the same generation) or intergenerational~(\acp{BH} are from different generations). However, this won't be the case for the chirp mass distribution, as intergenerational mergers will create intermediate peaks between the prominent ones. This is because chirp mass is a function of the two components~\citep{Cutler:1994ys}, \begin{equation}
    \mathcal{M} = \frac{(m_1\,m_2)^{3/5}}  {(m_1 + m_2)^{1/5}},
\end{equation} 
where the subscript identifies the two masses. The heavier \ac{BH}  is called primary~($m_1$) and the lighter \ac{BH} secondary~($m_2$).
Using the mass value for the location of peaks in the \ac{BH} mass distribution, we can calculate the location of peaks we expect to find in the \ac{BBH} chirp mass distribution. These values are listed in Table~\ref{tab:loc_peaks_mchirp}. Other combinations are possible but, due to mass segregation, the rates of \acp{BBH} with large mass asymmetry are expected to be suppressed. This is evident in the presence of intra-generational peaks. Moreover, this is supported by direct N-body simulations of star clusters \citep{Park2017}.
%As an example, the recently announced observations GW241011 and GW241110~\citep{o4a_special} have two \ac{BH} masses that are consistent with the first and second generations. In addition, these observations also show high spins. A third observation, GW240526~\citep{o4b_cat}, exhibits similar properties.
%s
\begin{table}[!t]
    \centering
    \renewcommand{\arraystretch}{1.3}
    \setlength{\tabcolsep}{6pt}
    \begin{tabular}{lccccccc}
        \hline\hline
        Hierarchy & 1G+1G & 1G+2G & 2G+2G & 2G+3G & 3G+3G & 3G+4G & 4G+4G \\
        \hline
        Chirp Mass Location & 7.5 & 10.2 & 14.2 & 19.4 & 27.1 & 37.0 & 51.4 \\
        \hline\hline
    \end{tabular}
    \caption{Expected location of peaks in the \ac{BBH} chirp mass distribution for various intra- and inter-generational mergers. All listed values are in solar masses. These values have been calculated using the expected location of peaks in the \ac{BH} mass distribution~\citep{2025arXiv251025579T}. In the hierarchical merger scenario, the peak locations in the \ac{BH} mass distribution are expected to follow a pattern. In this case, the first peak location is fixed at 8.6~(1G), and the subsequent ones at $8.6\times1.9=16.3$~(2G), $16.3\times1.9=31.0$~(3G), and $31.0\times1.9=59.0$~(4G).}
    \label{tab:loc_peaks_mchirp}
\end{table}

In this naive hierarchical picture, we have not considered aspects such as mass gain from accretion, the dependence of mass loss in \acp{GW} on the \ac{BBH} parameters, the retention dependence of a \ac{BBH} on its parameters for it to undergo further mergers or \ac{BH} masses that do not maintain a hierarchy after repeated mergers. Moreover, we chose the first-generation mass to maximise the visual overlap between the expected and inferred peak locations, and we did not perform a statistical analysis to determine it. We currently intend to assess how well these values align with the peaks in the inferred mass distribution. As we show in the following sections, they match quite well. For the remainder of this article, we will refer to our proposal as \emph{the hierarchical merger scenario} to distinguish it from other proposals involving hierarchical mergers.

\section{The Chirp Mass Distribution}
\label{sec:mchirp_distr}

\begin{figure*}
\includegraphics[width=0.98\textwidth]{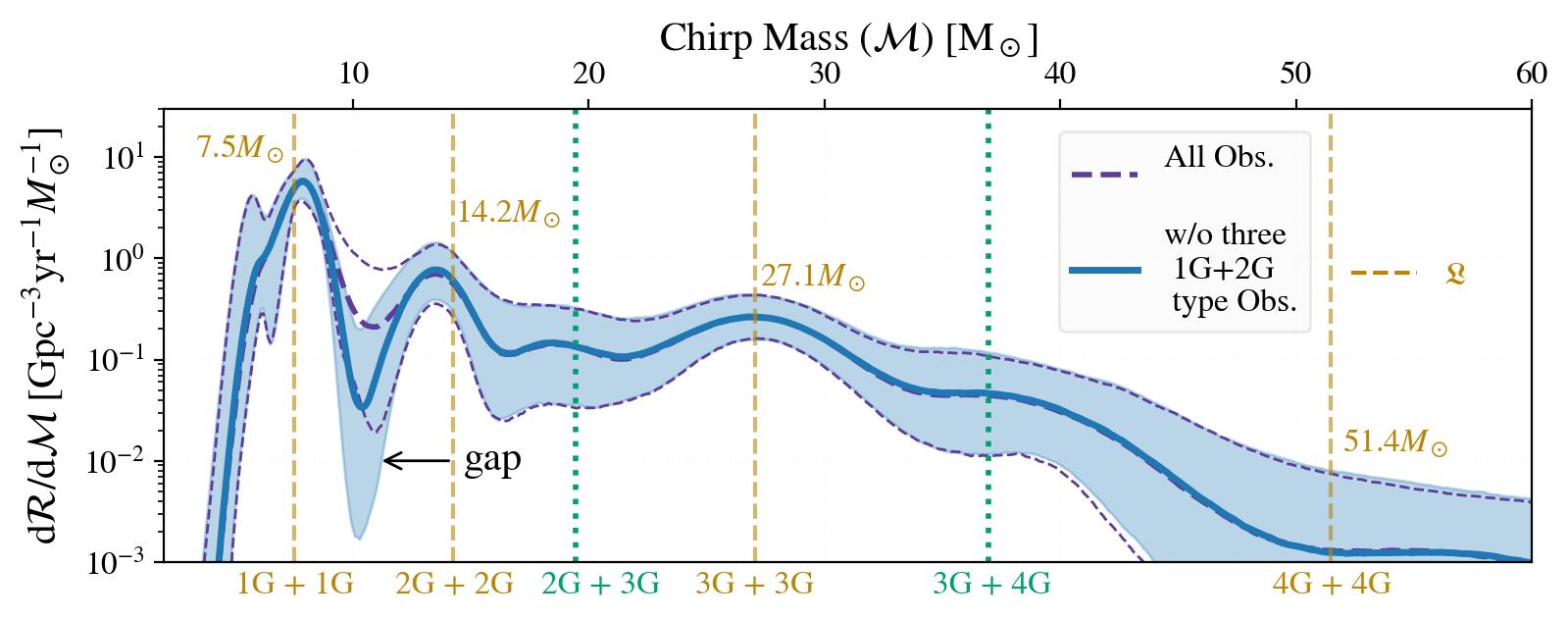}
\caption{The chirp mass distribution shows multiple peaks. Their locations match the expectations from the hierarchical merger scenario listed in Table~\ref{tab:loc_peaks_mchirp}. The \emph{mass ladder}, $\mathfrak{L}$, is a set of brown lines with a relative location that bears a factor of 1.9 from one to the next. These indicate the expected locations of intragenerational \acp{BBH}. The green lines indicate the locations of intergenerational \acp{BBH}. The dashed purple line represents the median of the inferred distribution using all observations along with the 90\% credible interval. The blue curve is the inferred median after excluding observations GW241011\_233834, GW241110\_124123~\citep{o4a_special} and GW240526\_093944~\citep{2026arXiv260527225T}. These observations exhibit large asymmetry and high spins. They belong to a group that several independent studies have investigated as 1G+2G mergers. Consequently, their chirp masses are expected to fill the gap~(or create a peak) between the first two peaks (please refer to Figure~\ref{fig:corr_mch_sz_q} and Section 4 in~\citealt{2021ApJ...913L..19T}). The blue band is the 90\% credible interval.}
\label{fig:mchirp_distr}
\end{figure*}

In this section, we report the chirp mass distribution inferred from GWTC-5. We used the mixture-model framework Vamana~\citep{2021CQGra..38o5007T} to perform this inference. We have detailed this analysis in Appendix~\ref{apndx:model}. Figure~\ref{fig:mchirp_distr} shows the inferred chirp mass distribution. It shows several peaks. These align with the expected chirp mass locations of various inter- and intra-generational mergers listed in Table~\ref{tab:loc_peaks_mchirp}. There is an emerging peak around 19$M_\odot$, which is consistent with a 2G+3G merger. The confidence in this peak is weak, and it is supported by 13 observations with mean chirp masses ranging from 17.5$M_\odot$ to 21.5$M_\odot$~\footnote{In comparison, there are 4 observations in the 15.5--17.5$M_\odot$ range, and 14 observations in the 21.5--23.5$M_\odot$ range. After correcting for selection effects, the excess appears around 19.5$M_\odot$}. However, 
this peak was anticipated earlier~(Section~4 in~\citealt{2022ApJ...928..155T}); while the statistical confidence of this individual feature is low, its appearance at the anticipated location is consistent with the predictive expectation of the hierarchical-merger interpretation. The appearance of this peak provides support to the peaks located around 14$M_\odot$ and 27$M_\odot$. 

 A Gaussian-like distribution, when added to itself, generates a new distribution with a mean and variance around twice those of the original distribution. The peak scales follow expectations from the hierarchical merger scenario, which, in essence, amounts to adding Gaussian-like distributions. The scales increase proportionally with generation; however, a more careful analysis is required to estimate them accurately. For example, aspects such as overlap between higher inter- and intra-generational peaks, larger measurement uncertainty for heavier masses, and the impact of the applied prior on Gaussian scales warrant investigation. 

 Moving to the heavier end of the mass distribution, we observe additional structure compared to our previous reports. This is because the maximum scale of the Gaussians that model the mass distribution depends on their locations. We have modified this proportionality in the version of Vamana presented here to better suit the increased number of observations. This is further discussed in Appendix~\ref{apndx:model}. However, due to the small number of heavy \acp{BBH}, the chirp mass region with $\mathcal{M} > 40$ has a large uncertainty. Here, we note that the mixture model infers the component mass plane. The structure in the chirp mass distribution is therefore not inferred directly.
\begin{figure*}
\includegraphics[width=0.98\textwidth]{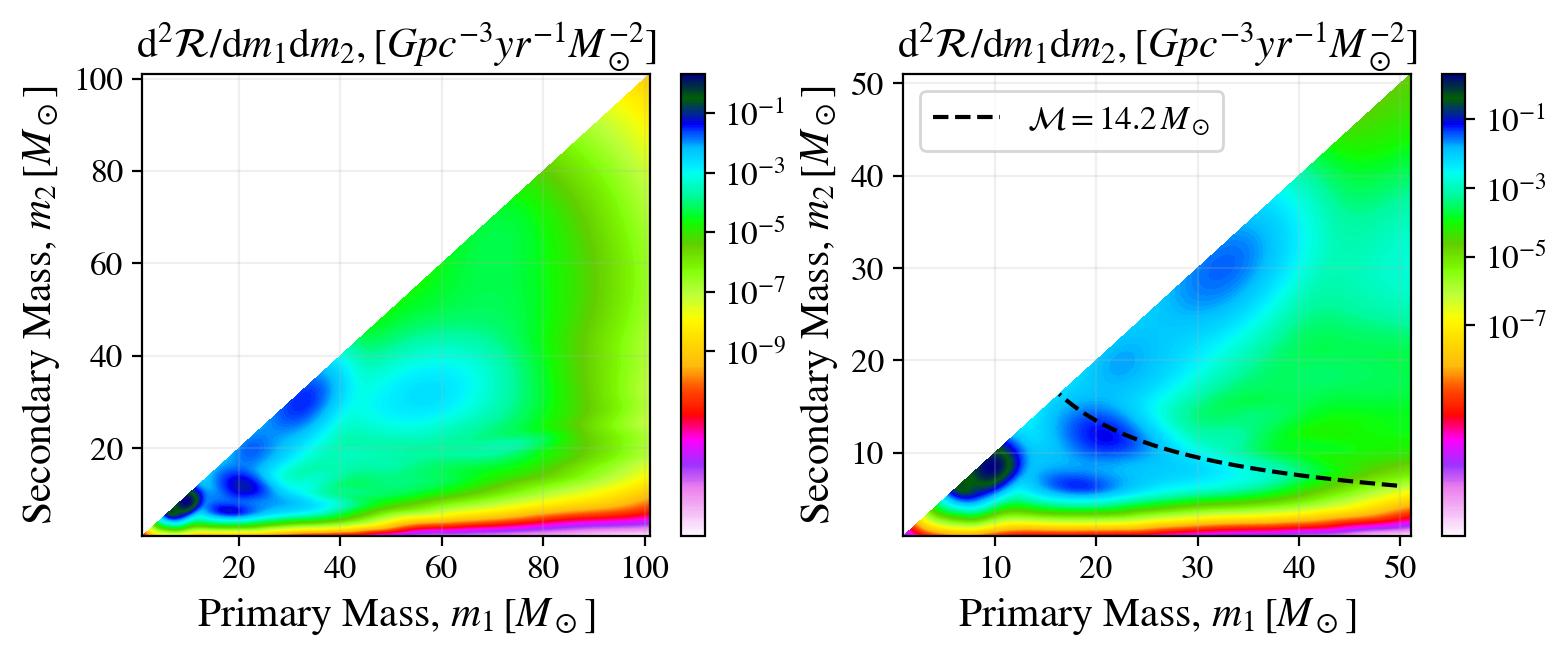}
\centering
\caption{Differential merger rate on the component mass plane in the local universe~(right plot is the lower-left quadrant of the left). Each overdensity present on this figure corresponds to a peak in the chirp mass distribution. The spin transition reported in several works occurs at two overdensities centred around 8--16$M_\odot$ and 30--60$M_\odot$. These have been investigated for two separate groups of 1G+2G mergers. The dashed track shows that the 14$M_\odot$ peak is prominent in the chirp mass; the two masses uniquely correlate to create this peak.}
\label{fig:corr_m1m2}
\end{figure*}
\begin{figure*}
\includegraphics[width=0.98\textwidth]{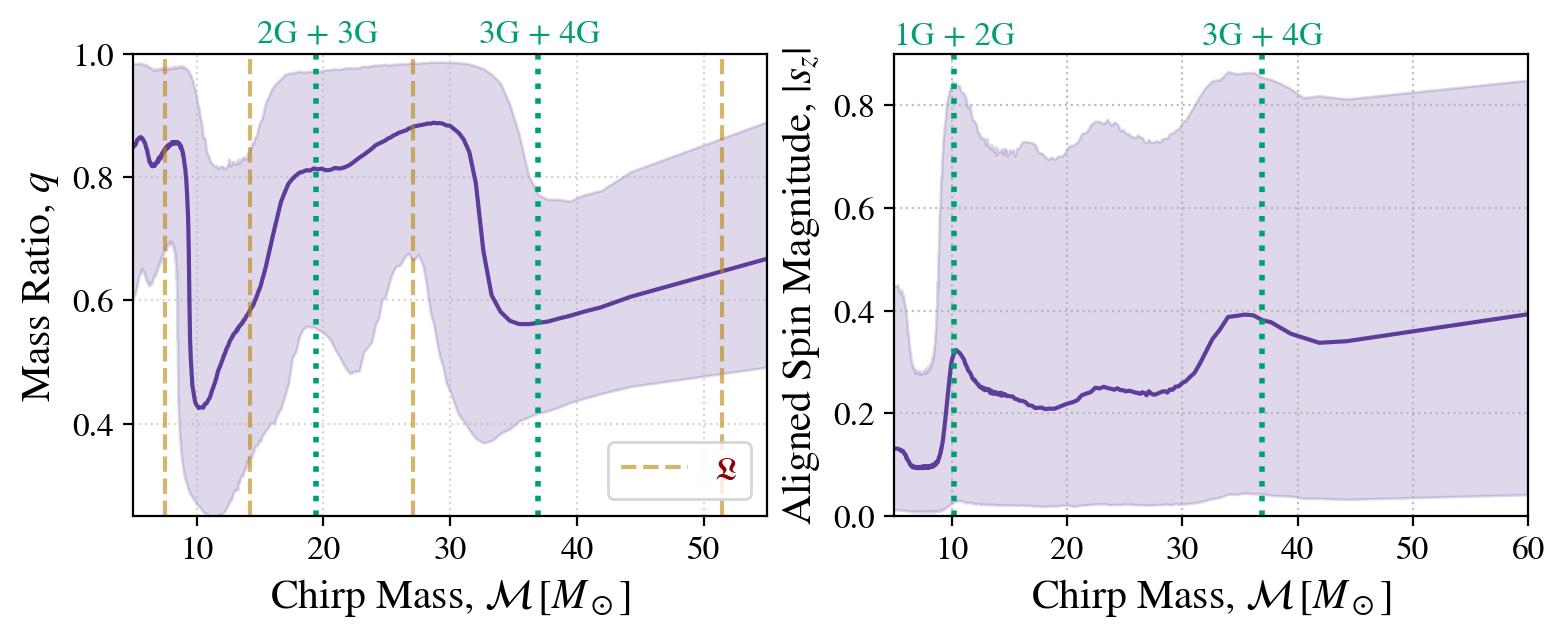}
\centering
\caption{Variation of mass ratio~(left) and aligned spin magnitude~(right) with the chirp mass. Several studies have reported spin transitions around chirp masses 10.2$M_\odot$ and 37$M_\odot$. In addition, the median mass ratio, shown by the solid curve, is approximately 0.5 at these chirp mass values. Some works have interpreted these transitions as two separate groups of 1G+2G hierarchical mergers; however, they naturally fit in the ladder structure shown in Figure~\ref{fig:mchirp_distr} as 1G+2G and 3G+4G mergers. The mass ratio varies substantially with chirp mass, which has motivated studies investigating this due to the contribution of several distinct subpopulations.}
\label{fig:corr_mch_sz_q}
\end{figure*}
\section{The High-Spin Binaries}
\label{sec:hispin}

Observations with primary masses around 10--20$M_\odot$ and $\geq 45M_\odot$ have recently gained attention. There is a transition in spins around these values, which has been suggested as a hint of subpopulations of high-spin \acp{BBH} consistent with separate groups of 1G+2G hierarchical mergers~\citep{2026arXiv260527226T}. The former spin transition is motivated by the characteristic mass ratio of observations such as GW241011\_233834 and GW241110\_124123~\citep{o4a_special} and GW240526\_093944~\citep{2026arXiv260527225T}. Formation of such \acp{BBH} requires a narrow peak in the mass distribution followed by a gap. These observations are expected to fill the gap between the first two intra-generational peaks. This is evidenced by the appearance of a local minimum in the inferred chirp mass distribution obtained after the removal of these observations and shown by the blue curve in Figure~\ref{fig:mchirp_distr}. This gap has been proposed to result from the supernova mechanism and has been investigated in multiple studies~\citep{2021ApJ...913L..19T, 2023ApJ...950L...9S, 2024ApJ...975..253A, 2025A&A...694A.186G, 2025ApJ...995..177T, 2025arXiv250820787W, 2026arXiv260401420L, 2026arXiv260525994G}. The latter spin transition coincides with a sharp decay in the secondary mass distribution, but not in the primary mass distribution. The decay in the secondary mass has been investigated in the context of a gap caused by the pair-instability supernova mechanism, and the absence of this decay in the primary mass distribution has been explained by a second group of 1G+2G hierarchical mergers. In this case, the first-generation \acp{BH} extend up to around 45$M_\odot$. 

We note that both of these spin transitions are part of the chirp-mass ladder. This is shown in Figure~\ref{fig:corr_m1m2} and~\ref{fig:corr_mch_sz_q}. The first transition happens around a chirp mass of 10$M_\odot$ and is consistent with 1G+2G mergers. We emphasise that this identification is not in tension with the 1G+2G interpretation of this group by independent works~\citep{2025arXiv250904637A, 2025arXiv250904151T}: both interpretations invoke hierarchical, dynamically assembled mergers. The second transition occurs around a chirp mass of 37$M_\odot$. Within the hierarchical merger scenario, and as previously noted~\citep{2025arXiv251025579T}, this group is naturally accommodated as a 3G+4G rung, in which the decay in the secondary mass near $\sim$45$M_\odot$ is consistent with the upper edge of the third-generation secondary population~(peaking near $\sim$31$M_\odot$) and the dearth before the fourth-generation scale~($\sim$59$M_\odot$)---an inter-rung feature rather than a pair-instability boundary. We do not adjudicate between the two interpretations here, but note that the ladder describes both spin transitions with a single mechanism, without a separately tuned stellar boundary at each location.

The high-spin observations are not limited to only these two mass ranges. In earlier work, we demonstrated that high-spin \acp{BBH} exhibit component masses and spin magnitudes consistent with hierarchical mergers (Section 4.1 of~\citealt{2025arXiv251025579T}). High-spin systems preferentially populate component-mass regions corresponding to peaks in the chirp-mass distribution. For instance, events such as GW241011\_233834 and GW241110\_124123 are consistent with a 1G+2G merger, while GW190412\_053044~\citep{2020PhRvD.102d3015A} is consistent with a 1G+3G configuration. These examples illustrate that hierarchical-merger signatures appear already in individual events. Crucially, the same fixed spacing that places GW241011\_233834 and GW241110\_124123 on the 1G+2G rung also predicts the intermediate 2G+3G peak near 19$M_\odot$ and the heavier rung discussed above, with no additional free parameters. The agreement on the light group is therefore not isolated but one anchor of a single ladder.

\begin{figure*}
\includegraphics[width=0.98\textwidth]{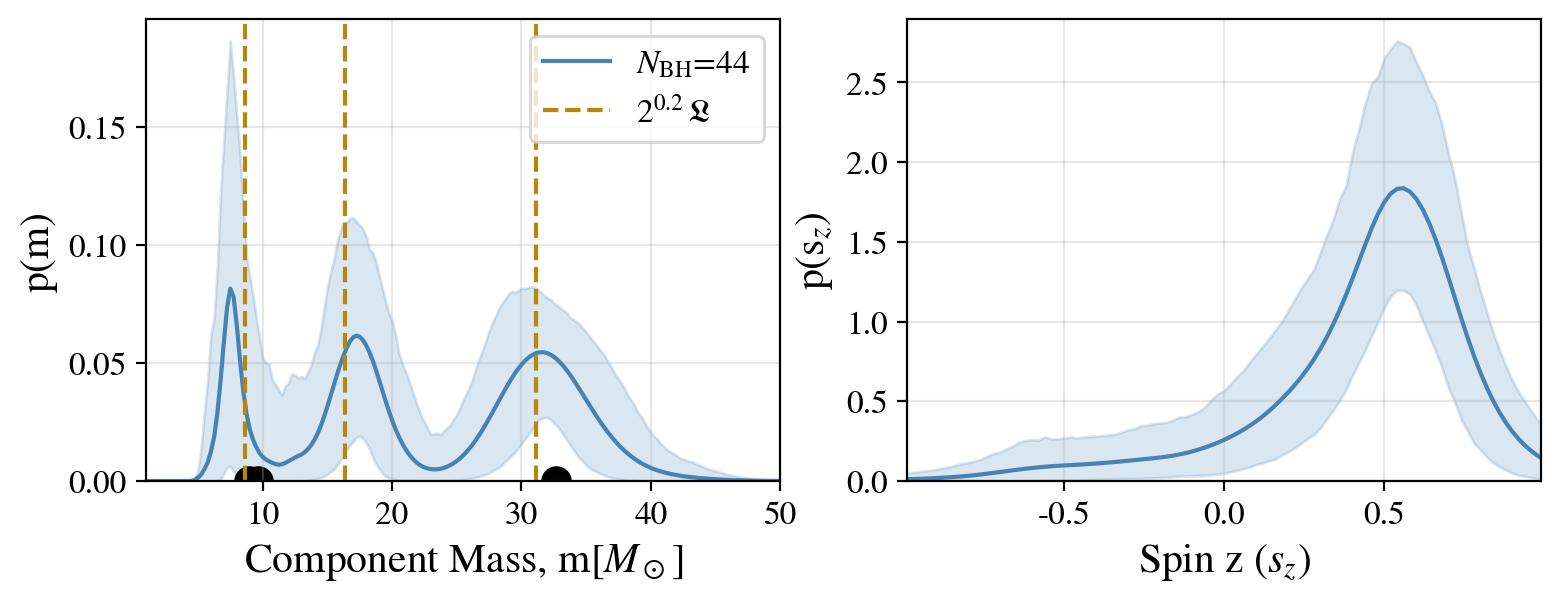}
\caption{The \ac{BH} mass~(left) and aligned spin~(right) distribution inferred from \acp{BBH} with $|\langle\chi_\mathrm{eff}\rangle| > 0.2$. The scaling factor between chirp mass and component mass of a comparable mass binary is $2^{0.2}$. This figure shows that multiple \acp{BBH} have mass ratios and spins consistent with the hierarchical merger scenario. The peaks in component mass match the chirp-mass ladder well after accounting for the scaling factor. The three black dots indicate the mass values of black holes found via astrometric observations with Gaia~\citep{2023MNRAS.518.1057E, 2023MNRAS.521.4323E, 2024A&A...686L...2G}; the forthcoming astrometric sample offers a test of the first-generation scale~(Section~\ref{sec:hispin}).}
\label{fig:highspin_pop}
\end{figure*}

The newly detected high-spin BBHs from the expanded GWTC‑5.0 catalog follow the same trend. An updated version of Fig. 12 in~\cite{2025arXiv251025579T}, shown in Figure~\ref{fig:highspin_pop} of this article, illustrates this behaviour: high-spin systems cluster near component-mass values that correspond to prominent chirp-mass peaks. We have provided further details in Appendix~\ref{apndx:hispin}.
While high-spin binaries trace hierarchical-merger expectations, they constitute only a fraction of the full BBH population. Nevertheless, the overall inferred chirp-mass distribution—after accounting for measurement uncertainties and selection effects—exhibits pronounced, regularly spaced structure that mirrors the locations traced by high-spin \acp{BBH}. The alignment of these component-mass features with chirp-mass peaks suggests that the hierarchical merger scenario may shape the mass distribution more broadly, even if mass-ratio and spin signatures are not uniformly apparent.

\section{Limited measurability of mass ratio and spins}
\label{sec:bias}
This juxtaposition raises an important question: if hierarchical mergers are responsible for the pronounced structure observed in chirp mass, why do the majority of BBHs not exhibit the mass ratios and spins typically associated with this formation channel~\footnote{\citealt{2025CQGra..42v5008K} discuss this tension in relation to 3G+3G peak.}? Is it due to the presence of contaminating channels~\citep{2025A&A...694A.186G}, due to the limited measurability of these parameters or both? This question is especially interesting as chirp mass is not inferred directly. As inference is made on the component mass plane, any model systematics can enter only through the unlikely route of correlation between masses.

We first examine the inferred distribution on the component-mass plane, which is shown in Figure~\ref{fig:corr_m1m2}. It exhibits several overdensities in two dimensions, each corresponding to a peak in the chirp mass distribution. Two of the overdensities are approximately centred around 8--16$M_\odot$ and 30--60$M_\odot$, and have the desired mass ratio and spin distribution expected from hierarchical mergers. Multiple works have investigated these as two separate groups of 1G+2G mergers. We note that these are part of the ladder structure in the chirp mass distribution and can be conveniently explained by the hierarchical merger scenario~(e.g. see~\citealt{2026arXiv260407456G} for the heavier group). However, the other overdensities do not prominently exhibit features consistent with hierarchical mergers. Most prominent among these mismatches is the peak around $14M_\odot$, which can be attributed to 2G+2G mergers. We have reported in several articles that this peak is substantially more prominent for chirp than for primary or secondary masses~\citep{2024MNRAS.527..298T, 2025ApJ...995..177T, 2025arXiv251025579T}. 

\begin{figure*}
\includegraphics[width=0.98\textwidth]{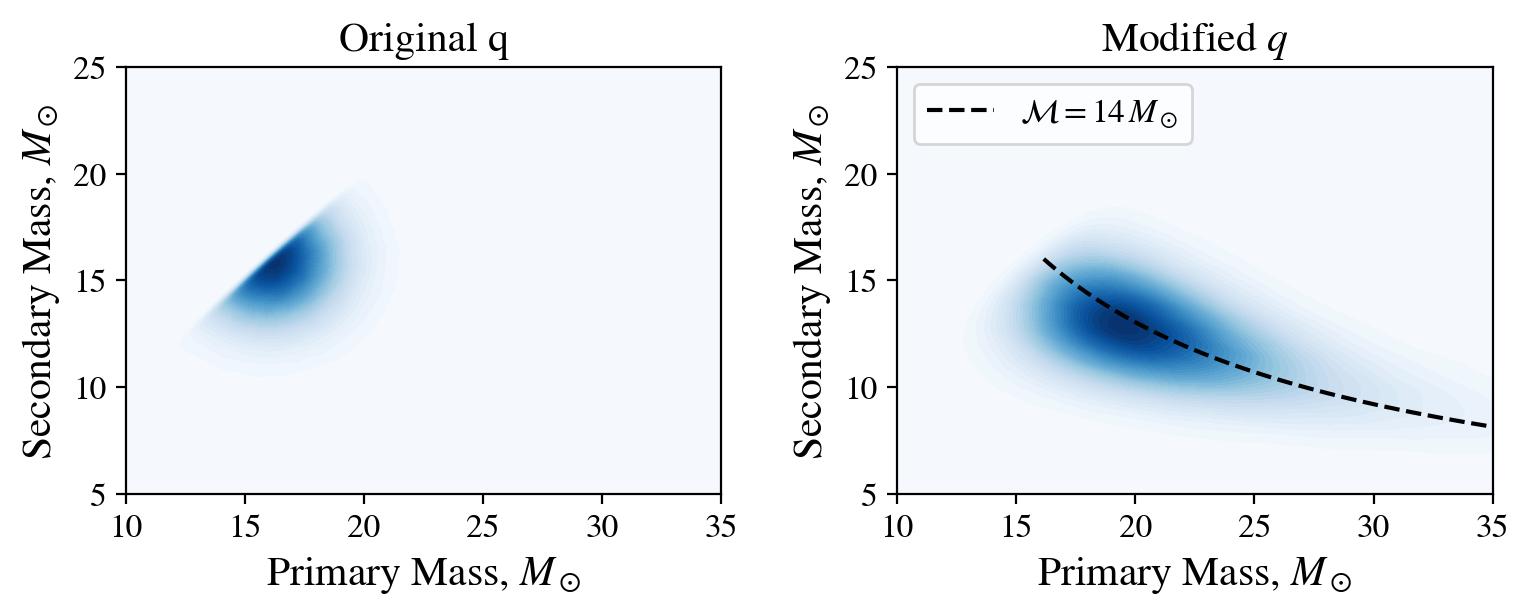}
\centering
\caption{An example illustrating the limited measurability of the mass ratio (and spins, as they are strongly correlated) compared to the chirp mass. A narrow peak around a chirp mass value of 14$M_\odot$ will be inferred as a localised overdensity~(left). However, the mass ratio may be susceptible to small systematic effects. If the mass ratio is modified, the inferred distribution still shows a peak in chirp mass, but the component masses now correlate uniquely~(right). Consequently, features in the primary or secondary mass may be less pronounced but more structured in the chirp mass distribution~\citep{2024MNRAS.527..298T}.}
\label{fig:unq_corr}
\end{figure*}

It seems plausible to attribute these mismatches to the limited measurability of mass ratio and spins. For example, focusing on observations with chirp masses near the 14$M_\odot$ peak, the ratio of the 90\% credible interval to the mean value for the measured chirp mass is approximately 4 times smaller than for the primary or secondary masses. Consequently, compared to other parameters, the inferred chirp mass is expected to be aligned to the corresponding astrophysical distribution most closely. Influenced by modelling, priors, selection effects, parameter degeneracies, parameter-dependent systematics, etc., the component masses are significantly more prone to systematic shifts than the chirp mass. This could manifest as a unique correlation between component masses, such that the chirp mass remains closest to the astrophysical distribution, while the mass ratio and spins undergo a systematic shift. We have further elucidated this in Figure~\ref{fig:unq_corr}. It shows a localised overdensity appears as a track on the component mass plane when chirp mass is close to the true value and mass ratio undergoes a systematic shift. The left panel shows the true distributions, and the right panel shows the measured mass ratio as normally distributed with mean/scale equal to 0.6/0.2. This choice is irrelevant, as it affects the distribution of rate along the track; the relevant aspect is the track's shape, which depends on the chirp mass's dependence on the component masses.

The accuracy of the chirp mass measurement reduces as the masses increase. Moreover, the mass ratio and spins are strongly correlated, leading to interdependence between measurements of these parameters~\citep{baird-2013, 2018ApJ...868..140T}. Consequently, the effect of any systematics on mass ratio or spin measurement may vary with the masses or other parameters. This may be the case, as seen in Figure~\ref{fig:corr_mch_sz_q}. Each feature in the chirp-mass distribution exhibits a different mass-ratio distribution. Although this conclusion is based on the median distribution, which has large credible intervals, it seems plausible. Here, we note that the variation in the mass-ratio distribution across mass range has been interpreted as evidence for distinct subpopulations~\citep{2026PhRvL.137b1403B, 2025ApJ...991...17R, Guttman:2026substructure}.
\section{conclusion}
\label{sec:conclude}
The release of GWTC-5.0 has expanded the catalog of \acp{BBH} observed through \acp{GW}, enabling a more precise resolution of the population’s mass distribution. Our analysis, based on 256 events from the catalog, confirms that the previously identified peaks in the chirp-mass distribution--located at approximately 7.5$M_\odot$, 14$M_\odot$, and 27$M_\odot$--remain prominent features. These peaks increase successively by a factor of roughly 1.9, which could be parsimoniously explained by the hierarchical merger scenario characterised by an approximate doubling of \ac{BH} masses across generations. Furthermore, the updated distribution reveals an emerging intermediate peak near 19$M_\odot$. The appearance of this feature at the location of 2G+3G intergenerational mergers underscores the predictive expectation of the hierarchical-merger scenario.

The hierarchical merger scenario also contextualises the behaviour of high-spin binaries within the population. Recent reports have proposed two separate groups of 1G+2G hierarchical mergers to explain observed spin transitions around primary masses of 10--20$M_\odot$ and $\geq 45M_\odot$. Crucially, we note that these proposed groups, located around chirp masses of 10$M_\odot$ and 37$M_\odot$, naturally align with 1G+2G and 3G+4G mergers, consistent with the structural expectations of the chirp-mass ladder. This is not merely a relabeling: the ladder unifies two transitions that are otherwise modelled as independent subpopulations under one spacing and predicts that the heavier group should exhibit two high-spin components rather than the spin asymmetry expected for first-plus-second-generation pairing. Independent astrometric mass measurements of detached black holes~(e.g.\ forthcoming Gaia data) may further probe the structure in the mass distribution. However, we also note that high-spin \acp{BBH} are not limited to these two specific groups of hierarchical mergers. Several observations, such as GW190412\_053044, exhibit component masses and spin signatures that support the extended ladder structure observed in the chirp-mass distribution, preferentially populating regions corresponding to the prominent peaks.

A challenge remains: aside from the 1G+3G and 3G+4G over-densities and a handful of observations, the broader population does not universally exhibit the mass ratios and spins typically associated with the hierarchical merger scenario. Among the possibilities, this mismatch may reflect contributions from other formation channels. Alternatively, this could reflect the limited measurability of these parameters. While the inferred mass ratio and spins are highly susceptible to parameter degeneracies, systematic shifts, and even astrophysical systematics, the chirp mass is measured with substantially higher accuracy and remains a much truer reflection of the underlying astrophysical distribution. Consequently, the distinct structural ladder observed in the chirp mass invites investigation into the role of hierarchical mergers as a fundamental driver of the \ac{BBH} mass distribution, even when component-level signatures may be obscured by measurement uncertainties.

The hierarchical merger scenario offers high scientific value. Because of its predictive power, it can shed light on the nature of stellar evolution and dynamics within star clusters. It also has a valuable use case in cosmology~\citep{2026arXiv260103257P, 2026arXiv260103347T, 2026arXiv260414290G}. 
\section*{Acknowledgments}
The author thanks Bernard Schutz, Alberto Vecchio and Thomas Dent for helpful discussions, and Shanika Galaudage for reviewing the manuscript.
This material is based upon work supported by NSF's LIGO Laboratory which is a major facility fully funded by the National Science Foundation. The author gratefully acknowledges computing resources provided by the LIGO laboratory, supported by the National Science Foundation grants, PHY-0757058 and PHY-0823459, and computing resources provided by Cardiff University, funded by Science and Technology Facilities Council grants, STFC grants ST/I006285/1 and ST/V005618/1.

\medskip
\noindent\textbf{Software:} \texttt{NumPy}~\cite{harris2020array}, \texttt{SciPy}~\cite{2020SciPy-NMeth}, \texttt{matplotlib}~\citep{Hunter:2007}, and \texttt{ASTROPY}~\citep{2022ApJ...935..167A}. Vamana can be installed by executing the command \texttt{pip install vamana}. Source code, additional files and results are available \href{https://github.com/vaibhavtewari/vamana}{\color{blue}here}.

\bibliographystyle{apsrev4-1}

% You should give the same name for your .bbl as your main .tex
% since it is a requirement for posting on ArXiv.
\bibliography{references}

\begin{appendix}

\section{Model Details}
\label{apndx:model}
We use the mixture model framework, Vamana~\citep{2021CQGra..38o5007T, 2025arXiv251025579T}, to infer the population. Vamana infers the two-component source--frame masses, aligned spins, and redshift distribution. Together, these provide a four-dimensional inference about the population. Table~\ref{tab:hyperparams} lists the hyperparameters. The binary parameters are estimated in the detector frame; we assume the Planck15 cosmology \citep{2016A&A...594A..13P} to convert to source-frame quantities. 

\begin{table*}[!t]
    \centering
    \renewcommand{\arraystretch}{1.3}
    \setlength{\tabcolsep}{6pt}
    \begin{tabular}{lccc}
        \hline\hline
        $\Lambda$ & Description / Modeled Parameter & Prior & Range \\
        \hline
        $w_i$ & Mixing weights, $w$ & Dirichlet($\bm{\alpha}$), $\alpha_{1\cdots N}= 1/N$ & 0--1 \\
        $\kappa_\mathrm{pop}$ & Populations's merger rate evolution index, $z$ & U / UL & $|\kappa_\mathrm{pop}| < 1.0$ / $1.0 < |\kappa_\mathrm{pop}| < 5.0$ \\
        $\kappa_\mathrm{comp}$ & Components's merger rate evolution index, $z$ & U & $|\kappa_\mathrm{pop} - \kappa_\mathrm{comp}| < 1.5$ \\
        $\mu^\chi_i$ & Gaussian's location, $s_z$ & U / UL & $|\mu^\chi_i| < 0.5$ / $0.5 < |\mu^\chi_i| < 0.9$ \\
        $\sigma^\chi_i$ & Gaussian's scale, $s_z$ & U & 0.5 / $\sqrt{N}$--1.5 / $\sqrt{N}$ \\
        $\mu_i^{m_1}$ & Gaussian's location, $m_1$ & U & 6$M_\odot$--75$M_\odot$ \\
        $\sigma_i^{m_1}$ & Gaussian's scale, $m_1$ & U & 0.155\,$\mu_i^{m_1}$/$\sqrt{N}$ -- 0.465\,$\mu_i^{m_1}$/$\sqrt{N}$ \\
        $\mu_i^{m_2}$ & Gaussian's location, $m_2$ & U & 6$M_\odot$--75$M_\odot$ \\
        $\sigma_i^{m_2}$ & Gaussian's scale, $m_2$ & U & 0.155\,$\mu_i^{m_2}$/$\sqrt{N}$ -- 0.465\,$\mu_i^{m_2}$/$\sqrt{N}$ \\
        $C_i^{m_1\text{--}m_2}$ & Covariance, $m_1$--$m_2$ & U & -0.75\,$\sigma_i^{m_1}\,\sigma_i^{m_2}$ -- 0.75\,$\sigma_i^{m_1}\,\sigma_i^{m_2}$ \\
        \hline\hline
    \end{tabular}
    \caption{Hyperparameters of the model used to infer the BBH population. U stands for Uniform, and UL for Uniform-in-log. $N$ is the number of components. Analysis imposes constraints, $\mu_i^{m_1} \geq \mu_i^{m_2}$ and $\mu_i^{m_2}/\mu_i^{m_1} \geq 0.2$ on the location of Gaussians. The merger rate of the population is modelled using a power law, $\mathcal{R} \propto (1 + z)^{\kappa_p}$. Gaussians are truncated to preserve a maximum aligned spin magnitude of one.}
    \label{tab:hyperparams}
\end{table*}

Several independent groups have reported new \ac{BBH} signals after analysing the open data from the LVK~\citep{Venumadhav:2019tad, Nitz:2018imz, Zackay:2019tzo, Venumadhav:2019lyq, Nitz:2020oeq, Nitz:2021uxj, Olsen:2022pin, 2023arXiv231206631W, Nitz:2021zwj, Kumar:2024bfe, Mishra:2024zzs, Mehta:2023zlk, Koloniari:2024kww}. However, since sensitivity estimates are required to accurately infer the population~\citep {2018CQGra..35n5009T}, we use only observations reported by the LVK collaborations~\citep{2019PhRvX...9c1040A, 2021PhRvX..11b1053A, 2024PhRvD.109b2001A, 2023PhRvX..13d1039A, 2025arXiv250818082T, 2026arXiv260527225T}. We select all observations with a false-alarm rate of at most one per year by at least one of the four search analyses used by the LVK. We also select only observations with a mean secondary mass greater than 3$M_\odot$. These criteria are satisfied by 256 observations. The estimated parameters for these \acp{BBH} and the injections used to correct selection effects are publicly available~\citep{zenodo_gwtc5, zenodo_gwtc5_injections_cumulative}. The GWTC-5.0 data release contained posterior samples obtained using a variety of waveform models. To remain consistent, we used posteriors obtained using {\sc{SEOBNRV(4}} or {\sc{{5)PHM}}}~\citep{2023PhRvD.108l4035P, 2023PhRvD.108l4037R} for all the observations.
\section{High Spin Population}
\label{apndx:hispin}
We perform an analysis identical to that in~\cite{2025arXiv251025579T} and refer the reader to that article for a description of the methodology. This analysis aims to find locations where the component masses of the high-spin binaries cluster. It uses a Gaussian mixture model with six components and infers three prominent locations. These locations are around 8.6$M_\odot$, 16.3$M_\odot$ and 31.0$M_\odot$. Our focus is on the \ac{BH} masses less than 40$M_\odot$. As individual spins are challenging to measure, we select \ac{BBH} with an effective spin magnitude greater than 0.2. We list all observations satisfying this criterion in Table~\ref{tab:mean_mass_hspin}.
\begin{table}[!t]
    \centering
    \renewcommand{\arraystretch}{1.3}
    \setlength{\tabcolsep}{6pt}
    \begin{tabular}{cc|cc|cc}
        \hline\hline
        Observation & Mean mass ($M_\odot$) & Observation & Mean mass ($M_\odot$) & Observation & Mean mass ($M_\odot$) \\
        \hline
        GW151226 & \textbf{15.5}--\textbf{7.5} & GW170729 & 53.1--\textbf{33.9} & GW190412\_053044 & \textbf{30.2}--\textbf{8.5} \\
        GW190513\_205428 & \textbf{35.6}--\textbf{19.5} & GW190517\_055101 & \textbf{38.3}--\textbf{25.3} & GW190620\_030421 & 58.3--\textbf{36.9} \\
        GW190719\_215514 & \textbf{37.7}--\textbf{21.3} & GW190805\_211137 & 47.8--\textbf{31.3} & GW190828\_063405 & \textbf{33.1}--\textbf{26.2} \\
        GW191103\_012549 & \textbf{12.6}--\textbf{7.8} & GW230624\_113103 & \textbf{30.9}--\textbf{16.1} & GW230820\_212515 & 63.9--\textbf{33.2} \\
        GW230825\_041334 & 48.0--\textbf{29.2} & GW230825\_041334 & 57.0--\textbf{31.0} & GW231113\_122623 & 41.5--\textbf{27.6} \\
        GW231118\_005626 & \textbf{19.5}--\textbf{11.3} & GW231223\_032836 & 47.1--\textbf{30.5} & GW231230\_170116 & 58.0--\textbf{34.9} \\
        GW240107\_013215 & 61.3--\textbf{29.7} & GW241011\_233834 & \textbf{18.9}--\textbf{6.2} & GW241110\_124123 & \textbf{16.8}--\textbf{8.3} \\
        GW240515\_005301 & \textbf{37.5}--\textbf{19.3} & GW240526\_093944 & \textbf{15.0}--\textbf{8.4} & GW240527\_183429 & 51.6--\textbf{33.8} \\
        GW240615\_113620 & \textbf{28.5}--\textbf{20.1} & GW240622\_004008 & \textbf{18.6}--\textbf{11.3} & GW241113\_163507 & \textbf{19.6}--\textbf{14.2} \\
        GW241116\_151753 & 57.2--\textbf{29.1} & & & & \\
        \hline\hline
    \end{tabular}
    \caption{The source frame, mean primary and secondary \ac{BH} masses for the high-spin BBHs. Although part of GWTC-5.0, this table does not include observations GW241011\_233834 and GW241110\_124123, as these observations were reported earlier~\citep{o4a_special} and were included in our previous report~\citep{2025arXiv251025579T}. The \ac{BH} masses, printed by the bold face, approximately cluster around 8.6, 16.3, and 31.0 $M_\odot$.}
    \label{tab:mean_mass_hspin}
\end{table}

\end{appendix}

\end{document}